\documentclass[pdflatex,sn-mathphys-num]{sn-jnl}

\usepackage{graphicx}%
\usepackage{multirow}%
\usepackage{amsmath,amssymb,amsfonts}%
\usepackage{amsthm}%
\usepackage{mathrsfs}%
\usepackage[title]{appendix}%
\usepackage{xcolor}%
\usepackage{textcomp}%
\usepackage{manyfoot}%
\usepackage{booktabs}%
\usepackage{algorithm}%
\usepackage{algorithmicx}%
\usepackage{algpseudocode}%
\usepackage{listings}%
\usepackage{amsmath}%
\usepackage{upgreek}
\usepackage{braket}
\usepackage{soul}

\usepackage{ulem}
\usepackage[binary-units=true]{siunitx}

\theoremstyle{thmstyleone}%
\theoremstyle{thmstyletwo}%

\theoremstyle{thmstylethree}%

\begin{document}

\title[Article Title]{
Unconventional superconductivity in locally non-centrosymmetric CeNi$_2$As$_2$
}


\author*[1,3]{\fnm{Felix} \sur{Morineau}}\email{felix.morineau@cpfs.mpg.de}\equalcont{These authors contributed equally to this work.}
\author[1]{\fnm{Jan} \sur{Knapp}}\equalcont{These authors contributed equally to this work.}
\author[1]{\fnm{Javier} \sur{Landaeta}}
\author[1]{\fnm{Thomas} \sur{L\"{u}hmann}}
\author[1]{\fnm{Lea} \sur{Richter}}
\author[1]{\fnm{Petra} \sur{Knappova}}
\author[1]{\fnm{Soumen} \sur{Ash}}
\author[1]{\fnm{Sushma Lakshmi} \sur{Ravi Sankar}}
\author[1,3]{\fnm{Konstantin} \sur{Semeniuk}}
\author[1,3]{\fnm{Elena} \sur{Hassinger}}
\author[1]{\fnm{Christoph} \sur{Geibel}}
\author[1]{\fnm{Manuel} \sur{Brando}}
\author[2]{\fnm{Daniel F.} \sur{Agterberg}}
\author[1]{\fnm{Andrew P.} \sur{Mackenzie}}
\author*[1]{\fnm{Seunghyun} \sur{Khim}}\email{seunghyun.khim@cpfs.mpg.de}

\affil[1]{\orgdiv{Max Planck Institute for Chemical Physics of Solids}, \orgaddress{\street{N\"othnitzer Stra{\ss}e 40}, \postcode{01187}, \city{Dresden}, \country{Germany}}}

\affil[2]{\orgdiv{Department of Physics}, \orgname{University of Wisconsin-Milwaukee},  \city{Milwaukee}, \postcode{53201}, \country{USA}}

\affil[3]{\orgdiv{Institute for Quantum Materials and Technology}, \orgname{Karlsruhe Institute of Technology}, \orgaddress{\street{Kaiserstra{\ss}e 12}, \city{Karlsruhe}, \postcode{76131}, \country{Germany}}}


\abstract{
Recent years have seen intense research on heavy fermion superconductivity, inspired by the discovery of multiple superconducting phases in UTe$_\text{2}$ and CeRh$_\text{2}$As$_\text{2}$.  
In the latter material, two superconducting phases observed under applied magnetic field oriented along the crystallographic $\textit{c}$ direction are associated with local inversion symmetry breaking.
To advance understanding of these phenomena, it is highly desirable to find further examples of heavy fermion superconductivity in materials with locally non-centrosymmetric structure.  
We have succeeded in this quest, by growing single crystals of the CaBe$_\text{2}$Ge$_\text{2}$ isomorph of CeNi$_\text{2}$As$_\text{2}$.
The resulting superconductivity brings more than we had foreseen.  
In addition to providing the opportunity to compare and contrast with that of CeRh$_\text{2}$As$_\text{2}$, the condensation occurs from an incoherent normal state quantitatively similar to that of UBe$_\text{13}$. 
Our findings therefore raise profound questions not just about superconductivity in the locally non-centrosymmetric structures, but about unconventional superconductivity itself.
}


\maketitle

\section{Main}\label{main}

Throughout the history of the research on correlated electron systems, material research has played a key role in advancing our understanding of unconventional superconductivity.
The fields of heavy fermion superconductivity \cite{steglich1979,Ott1983,Stewart1984_PRL,Stewart1984}, high temperature superconductivity \cite{Bednorz1986}, iron-based superconductivity \cite{Hosono2008} and graphene-based superconductivity \cite{Cao2018,Zhou2021} were all founded on the discovery of compounds with more or less unexpected properties.
In heavy fermion superconductivity, much of the focus in recent years has been on the multi-phase superconductivity seen in UTe$_2$ \cite{Ran2019,Braithwaite2019,Thomas2020}, CeRh$_2$As$_2$ \cite{khim2021}, and YbRh$_2$Si$_2$ \cite{Levitin2026}.
In CeRh$_2$As$_2$, the two-phase behaviour is thought to be associated with the presence of local centro-symmetry breaking in an overall centro-symmetric structure, and controlled by the dimensionless ratio of the characteristic energy for Rashba-type spin-orbit coupling and that for hopping between layers of Ce \cite{Maruyama2012,Fischer2023}.  
Interest in multiphase superconductivity has been broadened by recent observations on a dicalcogenide Ising superconductor \cite{Zhao2026}, highlighting the need to find further systems in which the underlying physics can be studied.  

The goal of the work that we report here was to search for another heavy fermion superconductors with the same crystal structure as CeRh$_2$As$_2$.  
We have succeeded in growing single crystals of the appropriate structural isomorph of CeNi$_2$As$_2$ for the first time, and have observed superconductivity.  
However, the results contain several surprises.  
Firstly, the superconductivity condenses from an incoherent normal state, with some similarities to that of UBe$_{13}$, the first time in over forty years that a second example of that phenomenon has been found in a heavy fermion material. 
Secondly, the temperature-magnetic field phase diagram invites a reexamination of some prominent assumptions that had been motivated by study of CeRh$_2$As$_2$ alone.

\begin{figure}[h]
\centering
\includegraphics[width=0.9\textwidth]{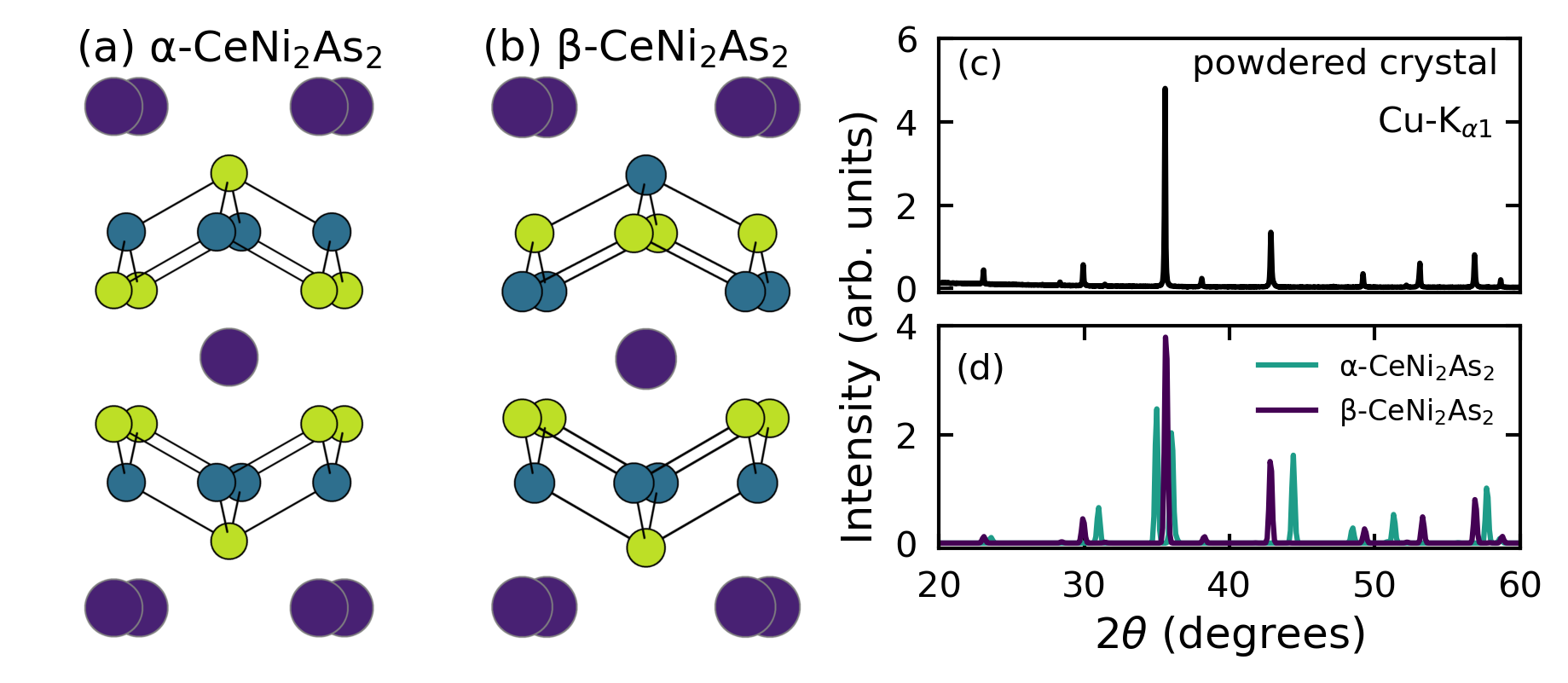}
\caption{
\textbf{Two isomorphic tetragonal structures of CeNi$_2$As$_2$.} \textbf {a,} $\upalpha$-CeNi$_2$As$_2$ in the body-centered ThCr$_2$Si$_2$-type and \textbf {b,} $\upbeta$-CeNi$_2$As$_2$ in the primitive CaBe$_2$Ge$_2$-type structure.
The major difference lies in the coordination of the constituting Ni-As layer that separates the Ce layer.
$\upalpha$-CeNi$_2$As$_2$ has Ni coordinated at the tetrahedral site surrounded with As while in $\upbeta$-CeNi$_2$As$_2$ one of the Ni-As layers has a complete interchange of the Ni and As positions. \textbf{c,} Powder x-ray diffraction pattern on crushed single-crystalline sample agrees with the expected pattern (\textbf{d}) for $\upbeta$-CeNi$_2$As$_2$ but clearly departs from the one for $\upalpha$-CeNi$_2$As$_2$. 
Detailed crystal structures are presented in Extended Data Tables \ref{tab:crystal_data}-\ref{tab:anisotropic_disp}.
}\label{fig1}
\end{figure}

Our choice of CeNi$_2$As$_2$ was based on evidence from solid-state chemistry studies that it can exist in both the $\upalpha$-CeNi$_2$As$_2$ isomorph with the common ThCr$_2$Si$_2$-type structure shared with, for example, CeCu$_2$Si$_2$ \cite{steglich1979}, URu$_2$Si$_2$ \cite{Mydosh2011}, BaFe$_2$As$_2$ \cite{Rotter2008}, and the $\upbeta$-CeNi$_2$As$_2$ isomorph which has the same CaBe$_2$Ge$_2$-type structure as CeRh$_2$As$_2$ \cite{khim2021}.  
Both are shown in Fig. \ref{fig1}a and b.  
$\upalpha$-CeNi$_2$As$_2$ has previously been grown in single crystal form and established to show a transition to antiferromagnetism of localized Ce moments at 4.8\,K, and a weak Kondo interaction \cite{Luo2012}. 
Although $\upbeta$-CeNi$_2$As$_2$ has been synthesized \cite{Ghadraoui1988,Suzuki2001}, low-temperature properties below 2 K were not examined.
For the present study, we succeeded in growing of $\upbeta$-CeNi$_2$As$_2$ from a Bi flux, as demonstrated by the x-ray diffraction data shown in Fig. \ref{fig1}c. 

\begin{figure}[h!]
\centering
\includegraphics[width=0.5\textwidth]{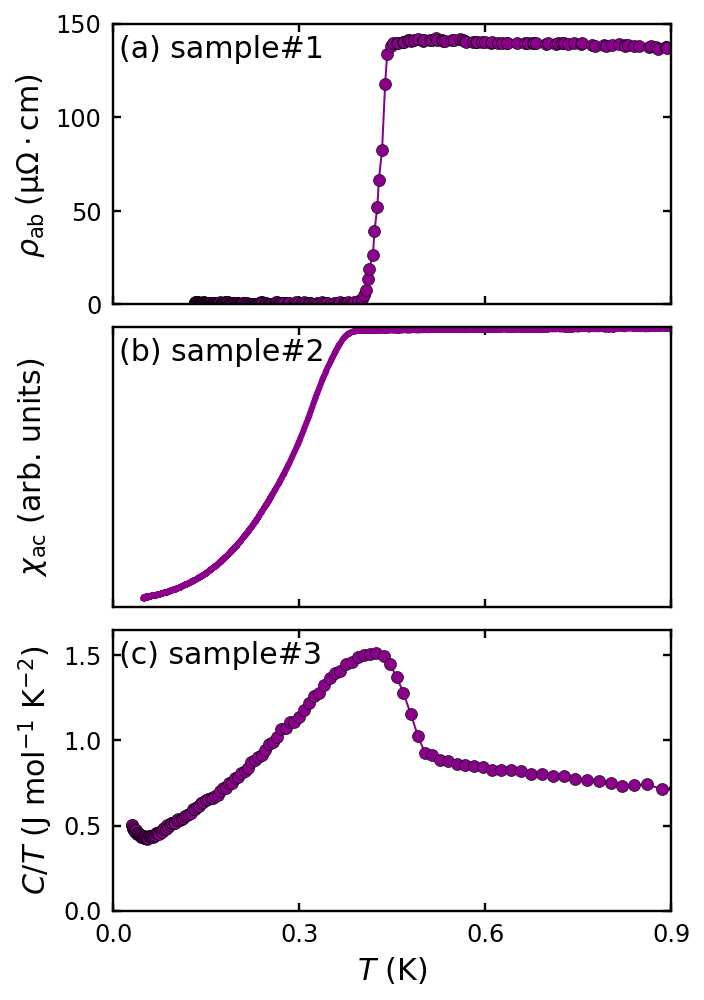}
\caption{
\textbf{Superconducting transition at 0.4\,K in $\upbeta$-CeNi$_2$As$_2$.}
\textbf {a}, Temperature dependence of $ab$-plane (bulk) resistivity, \textbf {b}, ac-magnetic susceptibility, and \textbf {c}, heat capacity divided by temperature $C$/$T$.
Sample\#1, \#2, and \#3 are representative: similar observations were made on a total of four crystals (see Extended Data Fig.\ref{FIGdiag} and Fig.\ref{FIGsusc}).
}\label{fig2}
\end{figure}

\begin{figure}[h]
\centering
\includegraphics[width=0.95\textwidth]{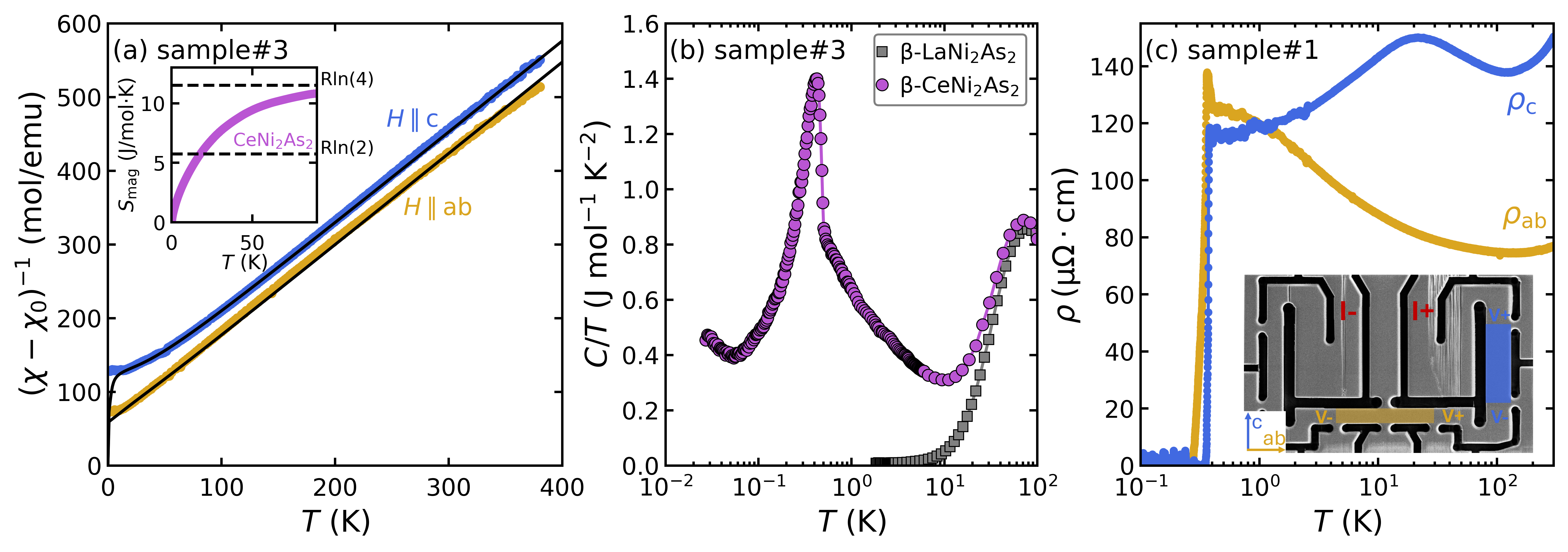}
\caption{
\textbf{Unusual normal state behaviors of the Kondo-lattice system $\upbeta$-CeNi$_2$As$_2$.} 
\textbf{a}, Inverse magnetic susceptibility after subtracting the temperature-independent term $\chi_0$. The solid lines are fits to the experimental data given by the crystal electric field (CEF) analysis (Extended Table \ref{CEF}), which determines the first-excited doublet to be located 10\,K above the ground state doublet. 
The inset shows magnetic entropy $S_\mathrm{mag}$($T$) of $\upbeta$-CeNi$_2$As$_2$. 
\textbf{b}, $C$/$T$ of $\upbeta$-CeNi$_2$As$_2$. No anomaly is seen at 4.8 K, confirming that the sample is free of any minority phase of $\upalpha$-CeNi$_2$As$_2$. Increasing $C$/$T$ with decreasing temperature below 10 K demonstrates a non-Fermi liquid behavior. The phonon heat capacity of $\upbeta$-CeNi$_2$As$_2$ is estimated from the heat capacity of $\upbeta$-LaNi$_2$As$_2$ in the same crystal structure, shown in the same panel. \textbf{c}, Temperature dependence of in-plane and $c$-axis resistivity, measured simultaneously in the FIB-sculpted device shown in the inset, cut from sample\#1. The small spike in $\rho_\mathrm{ab}$ at the onset of the superconducting transition is an artifact which does not repeat on measurements of $\rho_\mathrm{ab}$ on other samples (see for example the data in Fig. 2\textbf{a}).
}\label{fig3}
\end{figure}

As shown in Fig. \ref{fig2}, a superconducting transition is observed in single crystal $\upbeta$-CeNi$_2$As$_2$.  
In-plane resistivity, heat capacity and magnetic susceptibility all show the signatures expected of superconductivity at approximately 0.4\,K. 
The electronic heat capacity coefficient $\gamma$ (= $C$/$T$) is large, and comparison with that of LaNi$_2$As$_2$ (Fig. \ref{fig3}b) shows that the large value in $\upbeta$-CeNi$_2$As$_2$ originates from the presence of magnetic Ce.
A substantial fraction of this large $\gamma$ condenses into the superconducting state, indicating the formation of heavy fermion superconductivity.  
The $T_\mathrm{c}$ inferred from the heat capacity measurements is 30-50\,mK larger than that seen in the resistivity and susceptibility, which we attribute to sample-to-sample variation.  
The residual $\gamma$ (ignoring the nuclear Schottky term whose onset can be seen below 50\,mK) is approximately 0.3\,Jmol$^{-1}$K$^{-2}$, likely indicating that disorder plays a role in limiting the $T_\mathrm{c}$ of individual samples.
The resistivity at $T_\mathrm{c}$ is very large: 125\,\si{\micro\ohm}\,cm, and the resistivity rises between 1 K and $T_\mathrm{c}$.  
Similarly, $\gamma$ rises substantially for the same decrease in temperature, which already starts to increase below 10 K [Fig. \ref{fig3}(b)].
Both these properties indicate that the superconductivity condenses from a non-Fermi liquid normal state, and invite a further detailed examination.

In Fig. \ref{fig3}(a) we show the inverse $ab$-plane and $c$-axis magnetic susceptibility, derived from measurements of the magnetization, between 2\,K and 350\,K.
The observed Curie-Weiss behaviour yields an effective moment close to the 2.54 Bohr magneton per Ce expected of Ce$^{3+}$, and the magnetic anisotropy between the $ab$ plane and $c$ axis is weak.  
Further analysis of the magnetic susceptibility (see Extended Data Table \ref{CEF}) suggests the presence of two low-lying crystalline electric field (CEF) doublets with an extremely small split $\Delta_\textrm{CEF,1}$ of only $\sim$ 10\,K.  
The existence of the two doublets is reflected in the temperature dependence of the magnetic entropy $S_\textrm{mag}$ (inset to Fig. \ref{fig3}a) which increases without a plateau at $R$ln2 to approach $R$ln4 (where $R$ is the ideal gas constant).  
Analysis of $S_\textrm{mag}$ further suggests that the Kondo temperature is 12-36\,K, comparable to $\Delta_\textrm{CEF,1}$. 
In such circumstances, a quasi-quartet CEF quartet can emerge, since the Kondo coupling promotes mixing between the ground state and first-excited CEF doublets, which has been observed in CeRh$_2$As$_2$ \cite{Hafner2022}.  
As a result, unusual electronic quadrupole degrees of freedom that are usually forbidden in tetragonal Ce-systems can in principle be accessed.

The resistivity of $\upbeta$-CeNi$_2$As$_2$ is shown in Fig. \ref{fig3}c, for temperatures between 0.1 and 300 K.
Simultaneous measurement of $\rho_{ab}$ and $\rho_{c}$ was enabled by fabrication of the microstructure shown in the inset of Fig. \ref{fig3}c. Similar to the magnetization, the anisotropy of the resistivity in absolute terms is small: $\rho_{ab}$ and $\rho_{c}$ differ by no more than a factor of 2 at any temperature.  
The temperature dependences, however show distinct differences.  
$\rho_{c}$ exhibits a typical metallic Kondo-lattice temperature dependence, with a local maximum at around 20\,K and the decrease at lower temperatures as a coherent metallic state is formed from the hybridization of the localized 4$f$ electron with the itinerant conduction electrons.
The decrease of resistivity below the maximum is small, however, possibly indicating that the process of coherent state formation is incomplete.  
For $\rho_{ab}$, the situation is even more extreme: it looks as if the Kondo coherence maximum has not even been reached above $T_c$ and, as mentioned above, the superconductivity onsets at the remarkably high resistivity of $\rho_{ab}$ = 125\,\si{\micro\ohm}\,cm.  
We are aware of only one other heavy fermion superconductor in which the resistivity is so high at $T_\mathrm{c}$: the famous material UBe$_{13}$ \cite{Ott1983,Rauchschwalbe1986}.
There, a weak resistive maximum is seen at approximately 2\,K, but the coherent state has clearly not fully formed by the temperature at which the superconductivity condenses.
In the present case, condensation appears to happen from an even more incoherent normal state.  
Surprisingly, it has taken more than four decades to discover a second experimental example of this fascinating phenomenon, and the observation is particularly timely because condensation from an incoherent metal has emerged as a central theoretical focus in recent years\cite{hartnoll2018,Esterlis2026,Abanov2026}.
It has been known for a long time that incoherence generally suppresses unconventional superconductivity, but it is now understood that the incoherence can emerge from singular interactions that simultaneously drive non-Fermi liquid behaviour and enhance $T_\mathrm{c}$.
Often this is achieved by tuning a Fermi liquid towards a quantum critical state with a $T$-linear resistivity. 
In the present case, similar to the situation in UBe$_{13}$ but even more pronounced, the metallic state appears to be even away further from a Fermi liquid, providing the theoretical challenge of identifying the singular interaction that favours the formation of the superconducting state.  
This aspect of our observations is therefore of considerable significance, even without the Rashba-related insights provided by $\upbeta$-CeNi$_2$As$_2$, which we now describe.

\begin{figure}[h]
\centering
\includegraphics[width=0.9\textwidth]{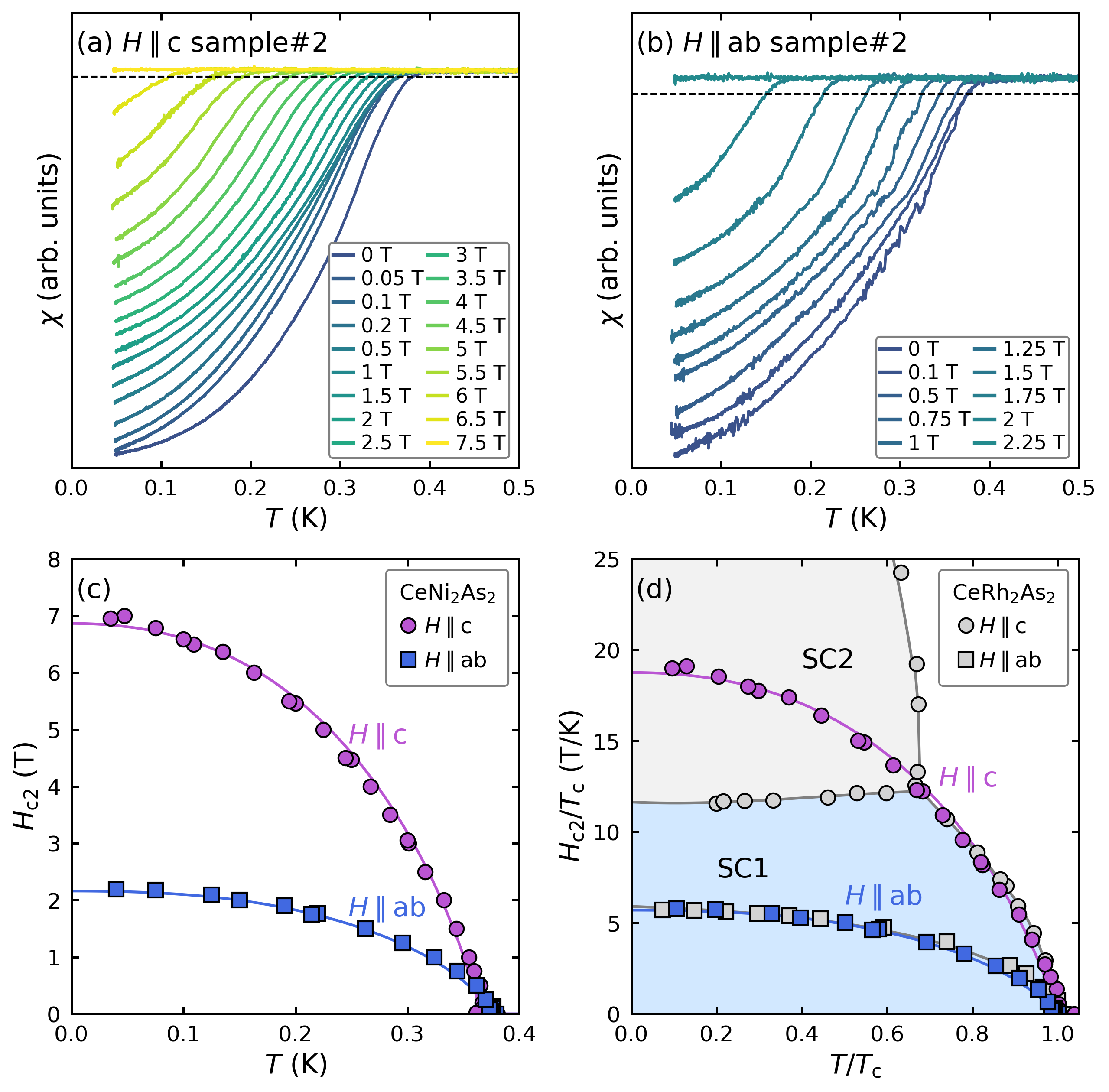}
\caption{
\textbf{Signature of Rashba superconductivity in $\upbeta$-CeNi$_2$As$_2$}
\textbf{a,b}, Temperature-dependent magnetic susceptibility for various applied fields parallel to the $c$ axis ($H$$\parallel$$c$) and $ab$ plane ($H$$\parallel$$ab$), respectively. The dashed horizontal lines denote the criterion used to determine $T_\mathrm{c}$ at each field.
\textbf{c}, The upper critical fields, $H_\mathrm{c2}$, obtained from the magnetic susceptibility data. The solid lines are fits to the experimental data using a theoretical model for the SC1 phase of CeRh$_2$As$_2$ \cite{khim2021}.
\textbf{d}, The normalized critical field, $H_\mathrm{c2}(T)$ divided by $T_\mathrm{c}$ ($H_\mathrm{c2}$/$T_\mathrm{c}$), as a function of reduced temperature $T$/$T_\mathrm{c}$. The corresponding normalized critical field (gray symbols) for CeRh$_2$As$_2$ ($H$$\parallel$$c$ and $H$$\parallel$$ab$ ([110]) are drawn for comparison. The shaded blue and gray regions refer the SC1 and SC2 phase of CeRh$_2$As$_2$, respectively. The gray solid lines are guides for the eye.}
\label{fig4}
\end{figure}

First, we turn to the superconducting critical fields which highlight the role of Rashba spin-orbit coupling (SOC). 
Specifically, by tracing the onset of the diamagnetic response in ac susceptibility under magnetic field (Figs. \ref{fig4}a and \ref{fig4}b), we establish the magnetic field - temperature phase diagrams shown in Fig. \ref{fig4}c.
Notably, in the zero temperature limit, upper critical fields are $H_{c2}$ = 2.2\,T and 6.9\,T for the $ab$-plane and $c$-axis fields respectively.
These $H_{c2}$ both exceed the Pauli-paramagnetic limit of $H_\mathrm{P} \approx$ 1.85$T_\mathrm{c} \approx$ 0.7 T. 
Often this is taken as an indication of spin-triplet superconductivity. We can further examine this possibility by estimating the orbital limiting critical field.
Using the slope of $dH_\mathrm{c2}/dT$ at $T_\mathrm{c}$ (-26 and -61 T/K for $ab$-plane and $c$-axis), we find that the $ab$-plane and $c$ axis orbital limiting fields are approximately (in the dirty limit) 7 and 16 T, respectively. 
Since both these fields are greater than the measured $H_\mathrm{c2}$, we rule out spin-triplet superconductivity which is typically orbital limited for some field orientation.
We further conclude that the critical field is Pauli limited but the effective Pauli-limiting field is enhanced.
Moderate violations of the Pauli limit are common among heavy-fermion superconductors and the original estimate of the Pauli limiting field can be modified by involving material-specific parameters such as the $g$-factor of the conduction electron, the effective mass, and the size of the superconducting gap \cite{Squire2023}.
However, in $\upbeta$-CeNi$_2$As$_2$, the in-plane $H_\mathrm{c2}$ is much more strongly Pauli suppressed than the $c$-axis $H_\mathrm{c2}$, implying that the modification of the Pauli limiting field is primarily given by a Rashba-like SOC \cite{Sigrist2009,Fischer2023}.
A similar anisotropy in $H_\mathrm{c2}$ has been observed in the non-centrosymmetric systems CeRhSi$_3$ and CePtSi$_3$ and in the locally non-centrosymmetric isostructural superconductor CeRh$_2$As$_2$. 
In all these cases, either a Rashba SOC or a local Rashba SOC is expected from symmetry considerations \cite{Sigrist2009,Fischer2023,khim2021}.

In $\upbeta$-CeNi$_2$As$_2$, there are two Ce per unit cell with each Ce belonging to a square lattice layer of Ce atoms. 
While the crystal has global inversion symmetry, this inversion symmetry relates these two square lattice layers.  
Since individual layer lacks inversion symmetry, it exhibits a layer-dependent Rashba SOC, here characterized by an energy scale $\alpha_\mathrm{R}$, that is of opposite sign on the two layers. 
In addition, there exists a hopping between these two layers, characterized by an energy scale $t_c$. 
Rashba superconductivity can appear when $\alpha_\mathrm{R}/t_\mathrm{c}$ is large, which can occur when the layers are well separated \cite{yoshida2012,Fischer2023}, or from non-symmorphic symmetries found in the CaBe$_2$Ge$_2$-type structure (the latter occurs when electrons have Bloch momenta near the Brillouin zone boundary \cite{Cavanagh2022}).
Using models for CeRh$_2$As$_2$ \cite{khim2021,Landaeta2022}, 
the SC parameters for $\upbeta$-CeNi$_2$As$_2$ yields $\alpha_\mathrm{R}/t_\mathrm{c} \sim$ 3.5, compared to 3.4 estimated for CeRh$_2$As$_2$ \cite{khim2021}.

In the simplest picture, the large values of their inferred $\alpha_\mathrm{R}/t_\mathrm{c}$ ratios suggests that the Rashba physics is important for both materials.
In that context the comparison shown in Fig. 4d is surprising.
The superconducting phase diagrams of the two materials are almost identical for the low-field phases, but the  high-field phase SC2 seen in CeRh$_2$As$_2$ for $H{\parallel}c$ is absent in $\upbeta$-CeNi$_2$As$_2$.
Within the commonly used description for the appearance of the high-field odd-parity state \cite{khim2021,Fischer2023,Landaeta2022}, the only two parameters that appear are $\alpha_\mathrm{R}/t_\mathrm{c}$ and $T_c^o/T_c^e$, where $T_c^o$($T_c^e$) is the transition temperature for an odd(even)-parity state. 
In principle, $T_c^o$ is independent from $\alpha_\mathrm{R}/t_\mathrm{c}$, however for $\textit{f}$-electron systems, as advocated by P.W. Anderson \cite{Anderson1985,Hazra2023}, site-local interactions are thought to drive superconductivity. 
In this case the ratio of pairing interactions for odd ($V_o$) and even-parity ($V_e$) states is  $V_o/V_e=\alpha_\mathrm{R}^2/(t_\mathrm{c}^2+\alpha_\mathrm{R}^2)$ \cite{khim2021}, where the even-parity (odd-parity) state is an in-phase (out of phase) layer superposition of a spin-singlet pairing state on an individual layer. 
This suggests that an odd-parity state should still appear for $\upbeta$-CeNi$_2$As$_2$.

The lack of an SC2 phase in $\upbeta$-CeNi$_2$As$_2$ indicates that the above conventional picture should be revisited. 
Specifically, the assumption that the origin of odd-parity superconductivity in CeRh$_2$As$_2$ is tied solely to the local non-centrosymmetric structure should be reconsidered, a conclusion that is also hinted at by recent experimental work on CeRh$_2$As$_2$ \cite{Semeniuk2024}. 
One generic consequence of this assumption is that $T_c^o<T_c^e$ but an examination of the evolution of the phase diagram of CeRh$_2$As$_2$ under pressure has revealed that the ratio of $T_c^o/T_c^e$ approaches 1 with increasing pressure and might surpass it if higher pressures can be achieved \cite{Semeniuk2024}. 
The lack of SC2 in $\upbeta$-CeNi$_2$As$_2$ further suggests that another mechanism is responsible for the odd-parity state. 
Indeed, some have already been suggested \cite{Moeckli2021,Lee2025,Nogaki2026}. 
Within the context of refs. \cite{Lee2025,Nogaki2026}  it is possible to understand why the odd-parity state might be suppressed in $\upbeta$-CeNi$_2$As$_2$.
Specifically the odd-parity states in these theories are tied to enhanced correlation effects driven by observed Van Hove singularities of the quasi-2D Fermi surface in CeRh$_2$As$_2$ \cite{Chen2024,Wu2024,Chen2024_prb}. 
The low temperature resistive anisotropy of $\upbeta$-CeNi$_2$As$_2$ is less than 15\%, compared with 200\% for CeRh$_2$As$_2$ \cite{Mishra2022}.
It is therefore possible that it is less 2D than that of CeRh$_2$As$_2$, suppressing the role of Van Hove singularities and hence the formation of the odd-parity state.
It would be of great interest to further examine the electronic structure and properties of $\upbeta$-CeNi$_2$As$_2$ to examine this proposal in more depth.  
It remains to be seen whether it satisfactorily accounts for the different phase diagrams, or whether other theoretical approaches are required.

It will also be important to examine the implications of the incompletely coherent normal state of $\upbeta$-CeNi$_2$As$_2$.  
To what extent is this the superconductivity of well-formed quasiparticles at all?  
We believe that our observations will re-energize discussions first stimulated by the properties of UBe$_{13}$ \cite{Cox1987,Ramirez1994}, and provide a useful new angle on the issue of the condensation of superconductivity from non-Fermi liquid normal states.  
Our observations on $\upbeta$-CeNi$_2$As$_2$ therefore challenge understanding of unconventional superconductivity at least two ways.

\section{Methods}\label{sec11}

\subsection{Crystal growth}\label{growth}
Single crystals of CeNi$_2$As$_2$ and LaNi$_2$As$_2$ in the CaBe$_2$Ge$_2$-type structure were grown by the Bi-flux method. 
Elemental metals with the molar ratio of Ce(La):Ni:As:Bi = 1:2:2:30 were placed in an alumina crucible which was subsequently sealed in a Ta tube under argon at a partial pressure of 900 mbar. 
The Ta tube was heated to 1100 $^{\circ}$C for 5 days, then slowly cooled down to 600 $^{\circ}$C over one week. 
Grown single crystals were extracted by centrifuging at 650 $^{\circ}$C.
Residual Bi on the surface of crystals was removed by diluted nitric acid.
The composition of crystals are characterized to be close to the nominal composition by energy dispersive x-ray spectroscopy (EDS) measurements.  

\subsection{X-ray diffraction}\label{diffraction}
Single crystal x-ray diffraction measurements were carried out by a Rigaku XtaLAB Mini II diffractometer with Mo-K$_{\upalpha}$ radiation ($\lambda$ = 0.71073 \text{\r{A}}).
The crystal structure was solved with SHELXT~\cite{Sheldrick2015} and then refined with SHELXL~\cite{Sheldrick2015_2}. Powder x-ray diffraction measurements data were obtained with Cu-K$_{{\upalpha}1}$ radiation ($\lambda$ = 1.5406 \text{\r{A}}) at room temperature.

\subsection{Fabrication of transport device}
\label{FIB_device}
The anisotropic transport measurements were performed on a microstructured device of $\upbeta$-CeNi$_{2}$As$_{2}$ sculpted using a plasma focused ion-beam (FIB). The procedure consisted in first milling a lamella of a single crystal of CeNi$_{2}$As$_{2}$ in the $ac$-plane, which would then be extracted and transferred on a thin suspended gold platform. The lamella was then milled in a shape that provides eight voltage probes along a unique current path, allowing measurement along two distinct $c$-axis channels and one $ab$-plane channel. Two additional voltage probes where added in the center of the ab channel for probing Hall effect. The dimensions ($l \times w \times h$) of the $c$-axis and $ab$-plane channels used in this paper are respectively 98.7\,\si{\micro\meter} $\times$ 32.4\,\si{\micro\meter} $\times$ 15.8\,\si{\micro\meter} and 124.6\,\si{\micro\meter} $\times$  20.0\,\si{\micro\meter} $\times$ 15.8\,\si{\micro\meter}. 
The microstructured device was obtained from a piece of the bulk of sample\#1. The resistivity measured along the $ab$-plane of the bulk sample and the microstructured device are almost identical to the exception of the presence a resistivity spike just before $T_{c}$ in the FIB device. That resistivity spike is a measurement artifact that can be associated to the inhomogeneity of the superconductive transition inside the microstructured channel. Some part of the material becomes superconductive between the voltage probes without connecting them which leading to a temporary elongated current path resulting in a higher effective resistivity.

\subsection{Specific heat and magnetic susceptibility}\label{specific_heat}
The specific heat was measured using the relaxation time method in a Quantum Design Physical Property Measurements System (PPMS, Quantum Design) down to temperatures of about 0.5\,K and a custom compensated heat-pulse calorimeter for temperatures between 0.04 and 4\,K and in magnetic fields up to 12\,T. 
DC magnetic susceptibility measurements were performed using a Magnetic Property Measurement System (MPMS-SQUID, Quantum Design) in a magnetic field of 0.2\,\si{\tesla}.
Ac-magnetic susceptibility was measured using a homemade set of compensated pick-up coils.
A superconducting modulation coil produced the excitation field of 175\,\si{\micro\tesla} at 5\,Hz.
The output signal of the pick-up coils was amplified using a low temperature transformer (LTT-m from CMR) with a winding ratio 1:100 and a low noise amplifier SR560 from Stanford.
A commercial susceptometer from CMR was also used.

\subsection{Resistivity}\label{Resistivity}
Resistivity down to 2 K was measured using a PPMS resistivity option with a standard four-point method using a current excitation of a 100\,\si{\micro\ampere}.
Below 2 K, measurement was carried out in a prototype adiabatic demagnetization refrigeration probe (designed by Solidcryo) utilized in a PPMS. In order to avoid too much heating at really low temperatures, the current excitation was lowered down to 10\,\si{\micro\ampere}.
Resistivity at low temperature under magnetic fields was measured in a dilution refrigerator also using a current excitation of 10\,\si{\micro\ampere}.
The signal was amplified by a low-temperature transformer with a winding ratio of 1:100 and the output of the transformer was measured using a SR830 lock-in amplifier at a frequency of 113.7\,Hz.

\backmatter





\bmhead{Acknowledgments}
The project was supported by the Max Planck Society.
S.K. and S.L.R.S. acknowledge support by the
Deutsche Forschungsgemeinschaft (DFG) - KH 387/1-1.
D.F.A. was supported by the Department of Energy, Office of Basic Energy Science, Division of Materials Sciences and Engineering under Award No DE-SC0021971.
K.S. and E.H. have received funding from the European Union’s Horizon Europe research and innovation programme under grant agreement No [Project 101125759 — Ixtreme].
J.K. and M.B. would like to thank the ERC for contributing to this research through JK's MSCA ID 101202931 ULT-SCES-NMR.
We thank Quantum Matter Group in Cavendish Laboratory (UK) for providing a customized demagnetization refrigeration (ADR) module. 
We also thank Marvin Klinger and Jorginho Villar Guerrero for their support and help with a prototype ADR probe from Solidcryo.com for mK electrical resistance measurements.
We thank Ulrich Burkhardt for EDX measurements and 
J\"{o}rg Schmalian and Gertrud Zwicknagl for useful discussions.

\section*{Declarations}


\begin{itemize}
\item Funding
Not applicable
\item Conflict of interest/Competing interests 
Not applicable
\item Ethics approval and consent to participate
Not applicable
\item Consent for publication
Not applicable
\item Data availability 
The datasets used in this study are publicly available in ** published sources. 
\item Materials availability
Not applicable
\item Code availability 
Not applicable
\item Author contribution
F.M. and J.K. contributed equally to this work.
S.K. and M.B. designed research and supervised the work.
S.K., S.A., and S.K.R.S. grew single crystals and characterized crystal structures.
F.M., K.S., L.R. fabricated FIB devices and measured resistivity.
J.K., P.K., F.M. and J.L. performed ac-magnetic susceptibility measurements. 
T.L. and M.B. carried out specific-heat measurements.
D.F.A. performed the theoretical calculations.
All authors contributed to the discussion of the results.
F.M., S.K, D.F.A. and A.P.M. wrote the paper with inputs from all authors.
\end{itemize}

\noindent

\newpage

\begin{appendices}

\section{Extended data}\label{secA1}

\begin{figure}[h]
\centering
\includegraphics[width=0.49\textwidth]{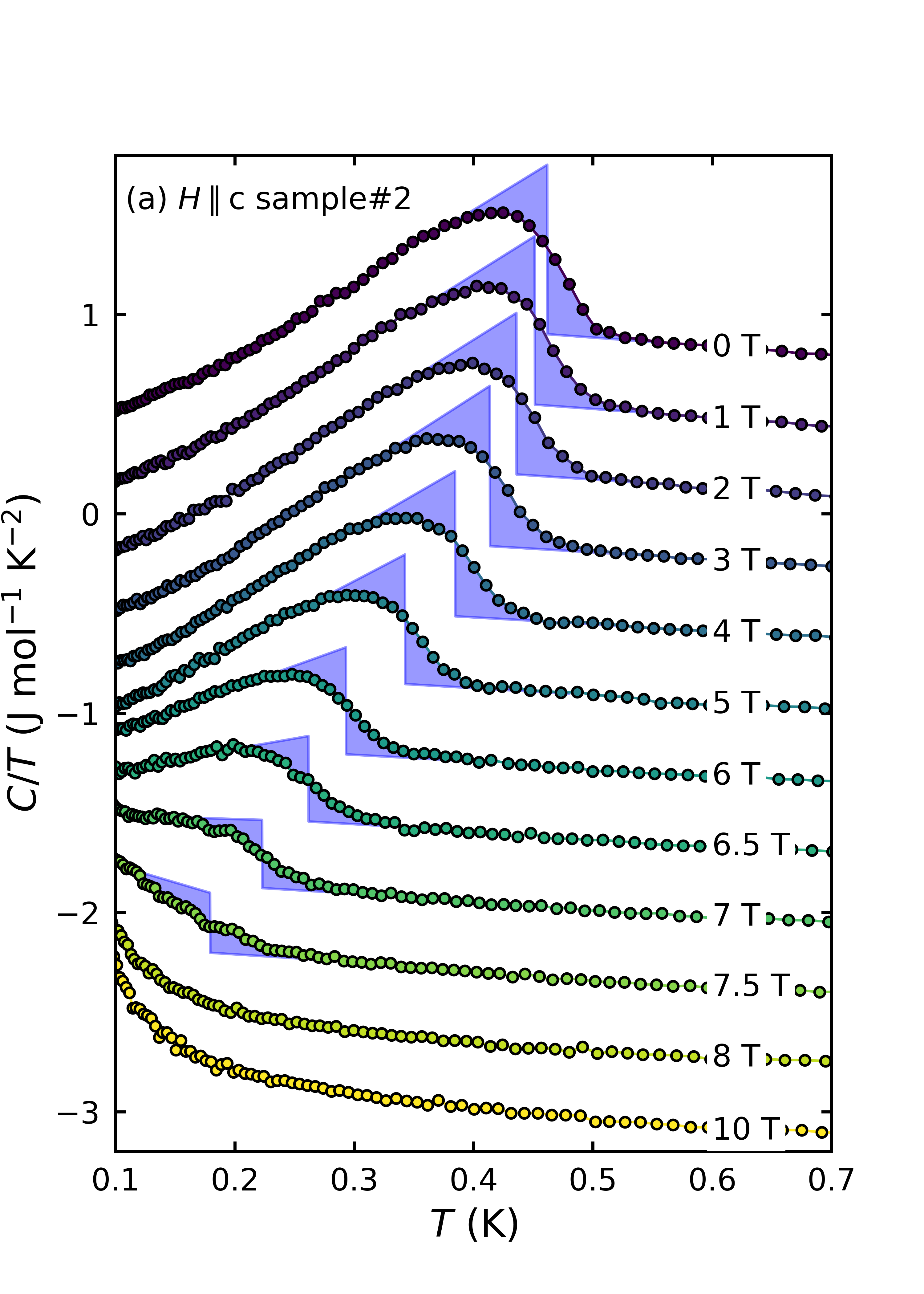}
\includegraphics[width=0.49\textwidth]{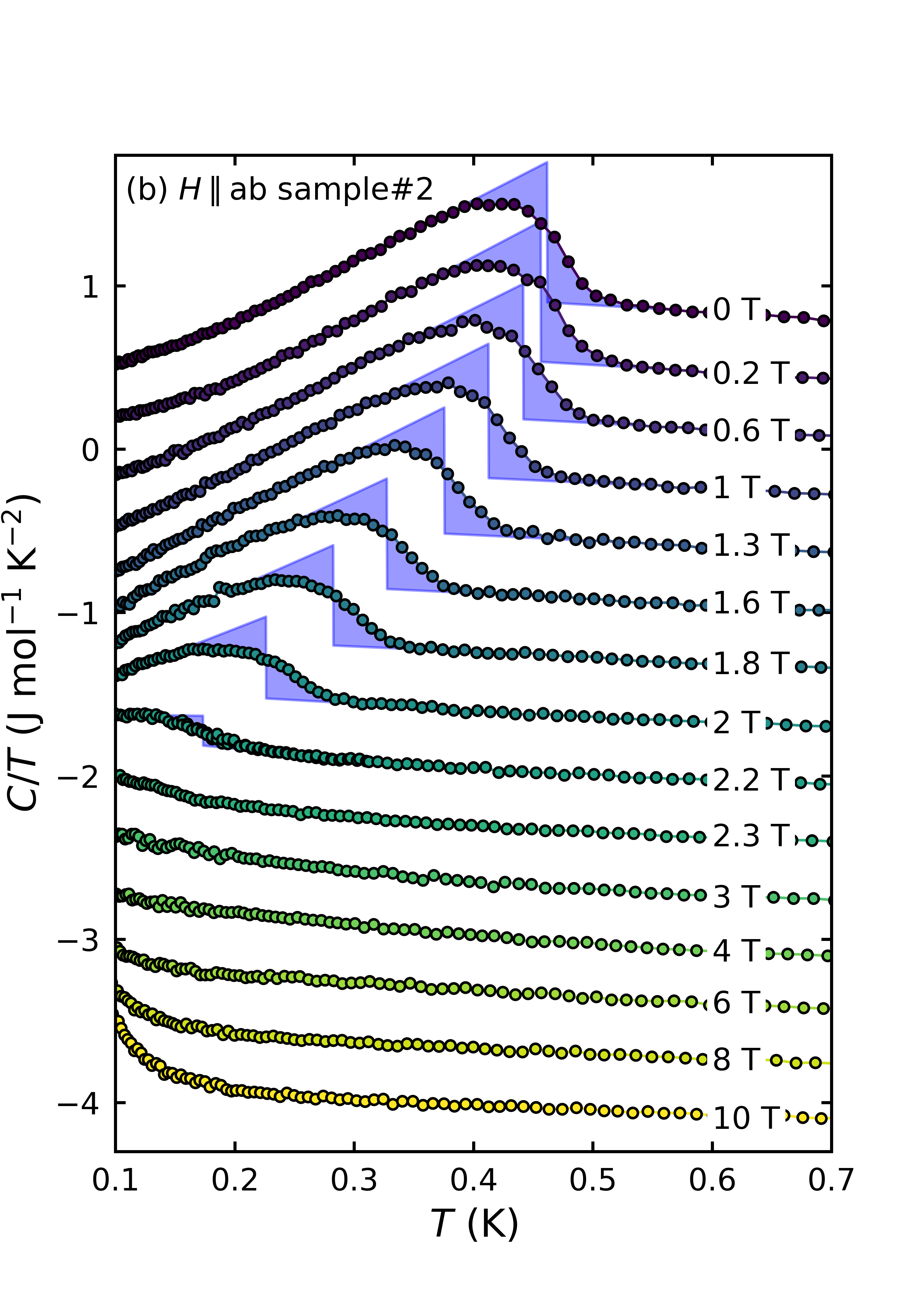}
\caption{
Temperature-dependent specific heat divided by temperature $C$/$T$ for (a) $H$$\parallel$$c$ and  (b) $H$$\parallel$$ab$. The shaded triangles denote the entropy balance below and above $T_\mathrm{c}$. The curves were shifted by -0.35\,Jmol$^{-1}$K$^{-2}$ for clarity.
}\label{equalS_cdir}
\end{figure}

\newpage

\begin{figure}[h]
\centering
\includegraphics[width=0.8\textwidth]{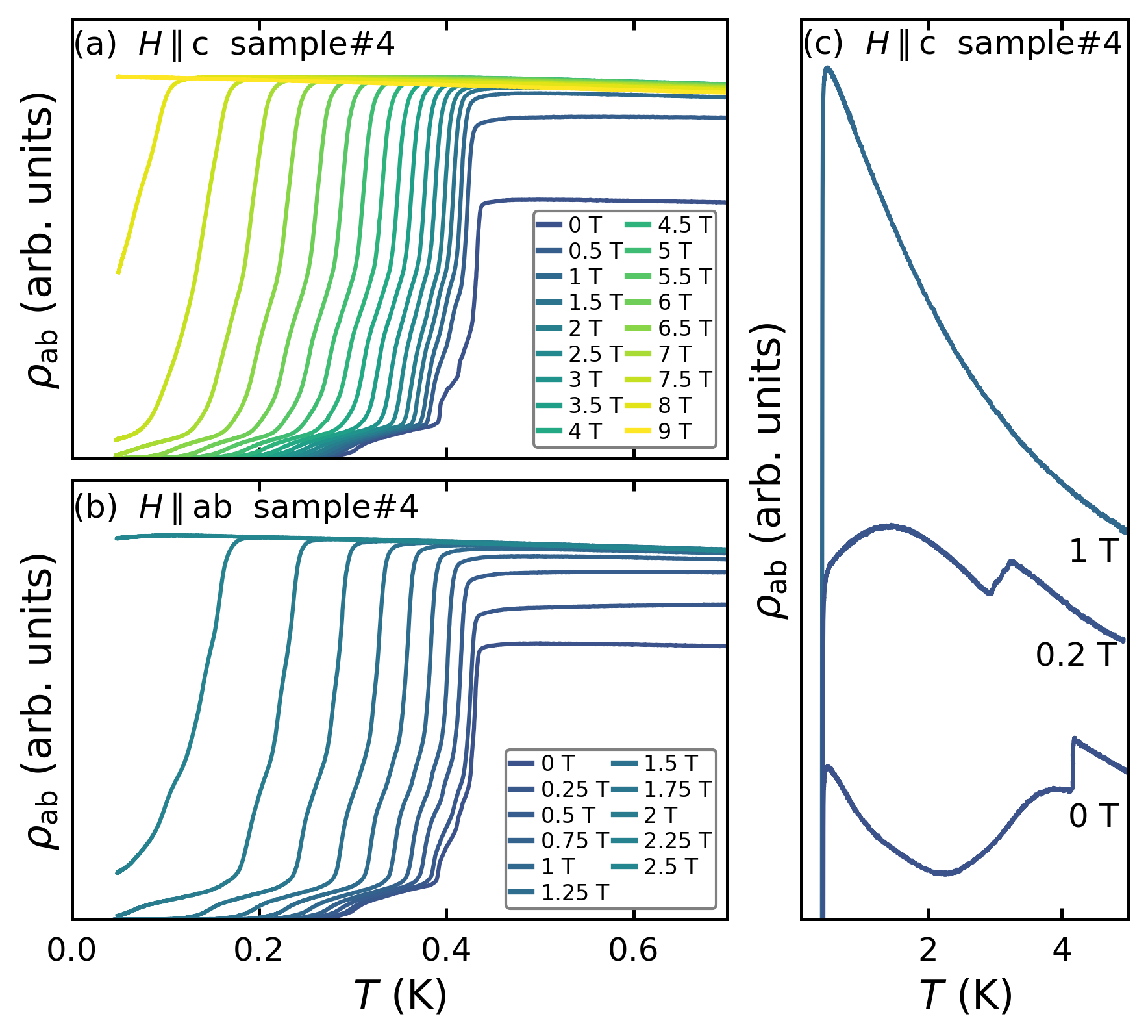}
\caption{
Temperature-dependent in-plane resistivity for (a) $H$$\parallel$$c$ and (b) $H$$\parallel$$ab$. The multiple steps observed in the SC transition indicates possible inhomogeneity of sample\#4. The strong change in resistivity under magnetic field above $T_\mathrm{c}$ can be attributed to the presence of a minor superconducting phase. (c) Resistivity in higher temperatures regions for applied field of 0, 0.2, and 1 T along the $c$-axis. The sharp resistivity drop observed at 4.1 K suggests the minor phase to be NiBi$_3$ \cite{Fujimori2000}.
}\label{resistivity_field}
\end{figure}

\newpage

\begin{figure}[h]
\centering
\includegraphics[width=1\textwidth]{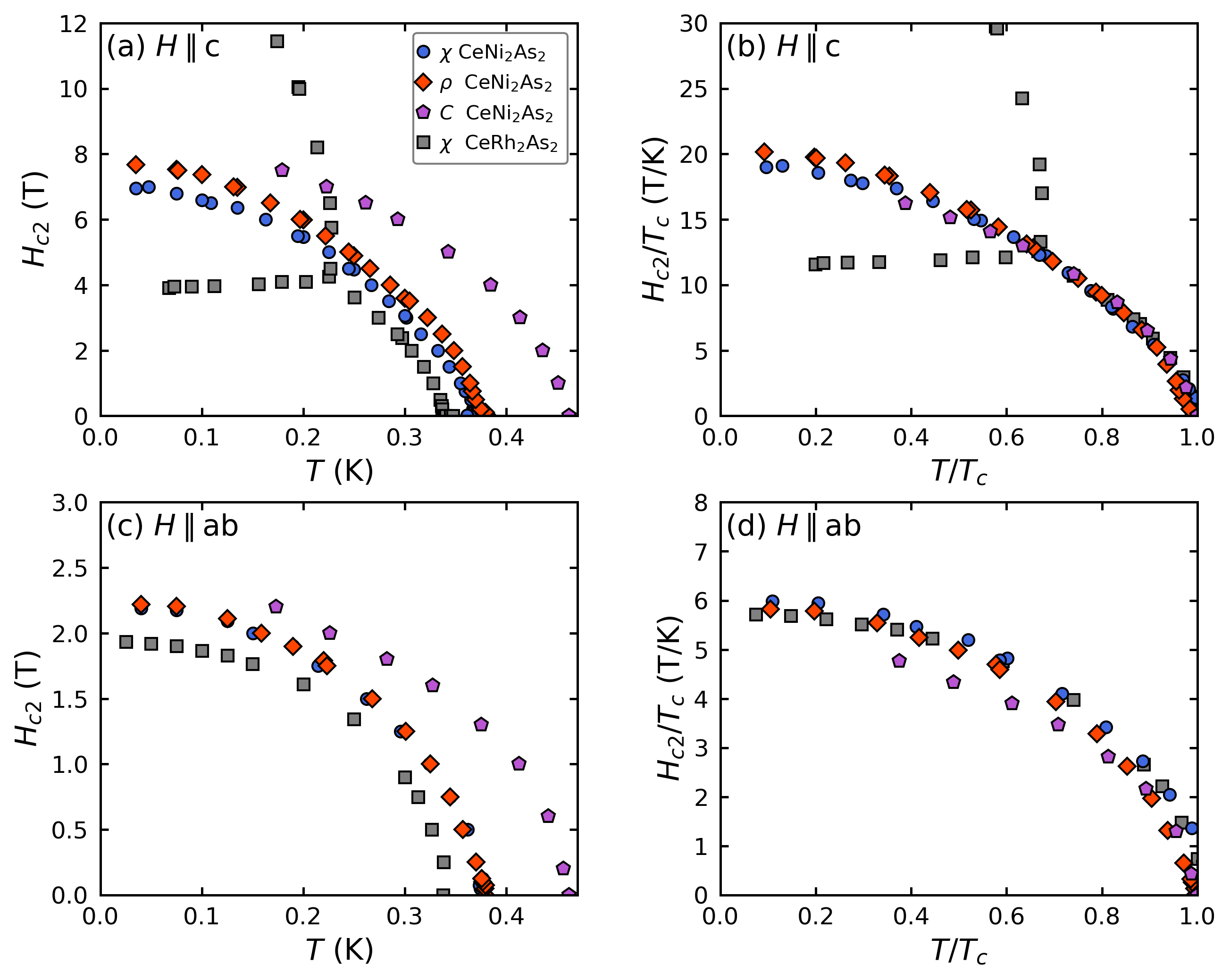}
\caption{
Summarized superconducting phase diagram $H_\mathrm{c2}$($T$) of $\upbeta$-CeNi$_2$As$_2$ determined from magnetic susceptibility, resistivity, and heat capacity for (a) $H$$\parallel$$c$ and (c) $H$$\parallel$$ab$, respectively.
$H_\mathrm{c2}$/$T_\mathrm{c}$ as a function of reduced temperature $T$/$T_\mathrm{c}$ for (b) $H$$\parallel$$c$ and (d) $H$$\parallel$$ab$, respectively. The corresponding normalized critical field for CeRh$_2$As$_2$ ($H$$\parallel$$c$ and $H$$\parallel$$ab$ ([110]) are drawn for comparison.
}\label{FIGdiag}
\end{figure}

\newpage

\begin{figure}[h]
\centering
\includegraphics[width=0.5\textwidth]{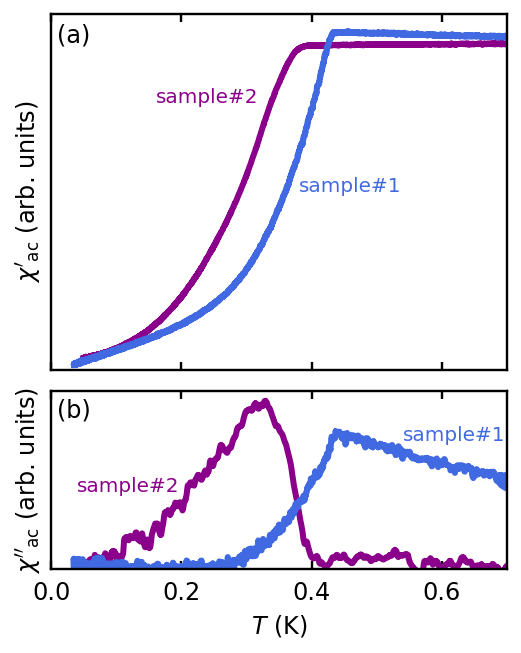}
\caption{Comparison of the complex zero-field ac-magnetic susceptibility $\chi_{\mathrm{ac}}=\chi'_{\mathrm{ac}}-i\chi''_{\mathrm{ac}}$ of sample\#1 and sample\#2 measured respectively using the commercial susceptometer from CMR and an homemade pick-up coils system. The real part $\chi'_{\mathrm{ac}}$ is shown in (a) and the imaginary part $\chi''_{\mathrm{ac}}$ in (b). Both measurements were scaled such that the size of their superconductive diamagnetic transitions are comparable.
}\label{FIGsusc}
\end{figure}

\newpage

\begin{table}[htbp]
\begin{tabular}{ccc} 
\hline
\hline
Wave function&Energy [K]\\
\hline
$\Gamma_{7}^{(2)}$ = 0.758$\ket{\mp\frac{3}{2}}$ - 0.651$\ket{\pm\frac{5}{2}}$  &107\\
$\Gamma_{7}^{(1)}$ = 0.758$\ket{\mp\frac{5}{2}}$ + 0.651$\ket{\pm\frac{3}{2}}$  &11\\
 $\Gamma_{6}$ = $\ket{\pm\frac{1}{2}}$ &0\\
\hline
\hline
CEF Parameters& $B_{2}^{0}$ = 2.29 K, $B_{4}^{0}$ = - 0.18 K,\\ 
&${\mid}B_{4}^{4}\mid$ =1.77 K, \\
&$\lambda_\textrm{MF}$ = 58 mol/emu\\ 
\hline
\hline
    \end{tabular}
      \caption{
The crystal field Hamiltonian is given by $H$ = $B_{2}^{0}O_{2}^{0}$ + $B_{4}^{0}O_{4}^{0}$ + $B_{4}^{4}O_{4}^{4}$ where $B_{n}^{m}$ are the crystal electric field parameters and $O_{i}^{j}$ are Stevens operators \cite{HUTCHINGS1964,Stevens1952}. 
The magnetic susceptibility is calculated from the CEF Hamiltonian by on the formula described in Ref.\cite{khim2021} based on the energy levels given above.
$B_{2}^{0}$ = 2.29 K is calculated from the relation, $B_2^0 = (\theta_\mathrm{ab} - \theta_\mathrm{c})\cdot\frac{10k_B}{3(2J-1)(2J+3)}$, where $k_\mathrm{B}$ is the Boltzmann constant, $\theta_\mathrm{ab}$ = 45.3\,K, $\theta_\mathrm{c}$ = 63.8\,K for $J$ = 5/2. 
$B_{4}^{0}$ = -0.17\,K, and $B_{4}^{4}$ = $\mid$1.77$\mid$\,K are obtained from the diagonalisation of the Hamiltonian.
}
\label{CEF}
\end{table}

\newpage

\begin{table}[htbp]
\centering
\caption{
Crystal structure refinement results for $\upbeta$-CeNi$_2$As$_2$.
}
\label{tab:crystal_data}
\renewcommand{\arraystretch}{1.1}
\begin{tabular}{l l}
\toprule
Chemical formula & CeNi$_2$As$_2$ \\
Temperature/K & 293(2) \\
Crystal system & tetragonal \\
Space group & P4/nmm \\
$a$ / \AA & 4.18450(10) \\
$b$ / \AA & 4.18450(10) \\
$c$ / \AA & 9.4378(3) \\
$\alpha$ / $^\circ$ & 90 \\
$\beta$ / $^\circ$ & 90 \\
$\gamma$ / $^\circ$ & 90 \\
$V$ / \AA$^3$ & 165.256(9) \\
$Z$ & 2 \\
$\rho_{\text{calc}}$ / g\,cm$^{-3}$ & 8.187 \\
$\mu$ / mm$^{-1}$ & 44.392 \\
$F(000)$ & 360.0 \\
Crystal size / mm$^3$ & 0.1 $\times$ 0.09 $\times$ 0.04 \\
Radiation & Mo K$\upalpha$ ($\lambda$ = 0.71073) \\
2$\theta$ range / $^\circ$ & 8.638 to 61.556 \\
Index ranges & $-5 \leq h \leq 6$, $-6 \leq k \leq 5$, $-13 \leq l \leq 13$ \\
Reflections collected & 11489 \\
Independent reflections & 188 [$R_{\text{int}}$ = 0.0465, $R_{\sigma}$ = 0.0096] \\
Data/restraints/parameters & 188/0/15 \\
Goodness-of-fit on $F^2$ & 1.210 \\
Final $R$ indexes [$I > 2\sigma(I)$] & $R_1$ = 0.0205, $wR_2$ = 0.0533 \\
Final $R$ indexes [all data] & $R_1$ = 0.0215, $wR_2$ = 0.0536 \\
Largest diff. peak/hole / e\,\AA$^{-3}$ & 1.05 / -1.10 \\
\bottomrule
\end{tabular}
\end{table}

\newpage

\begin{table}[htbp]
\centering
\caption{Fractional atomic coordinates ($\times 10^4$) and equivalent isotropic displacement parameters ($\mathrm{\AA}^2 \times 10^3$) for $\upbeta$-CeNi$_2$As$_2$. $U_{\text{eq}}$ is defined as $1/3$ of the trace of the orthogonalised $U_{ij}$ tensor.}
\label{tab:atomic_coords}
\renewcommand{\arraystretch}{1.15}
\begin{tabular}{l c c c c}
\toprule
Atom & $x$ & $y$ & $z$ & $U_{\text{eq}}$ \\
\midrule
Ce01 & 2500 & 12500 & 7484.4(4) & 8.2(2) \\
As02 & 2500 & 2500 & 3697.4(9) & 8.8(3) \\
As03 & 2500 & -2500 & 0 & 16.0(3) \\
Ni04 & 2500 & 7500 & 5000 & 13.2(3) \\
Ni05 & 2500 & 2500 & 1174.4(13) & 14.2(3) \\
\bottomrule
\end{tabular}
\end{table}

\begin{table}[htbp]
\centering
\caption{Anisotropic displacement parameters ($\mathrm{\AA}^2 \times 10^3$) for $\upbeta$-CeNi$_2$As$_2$. The anisotropic displacement factor exponent takes the form: $-2\pi^2[h^2 a^{*2} U_{11} + 2hka^*b^* U_{12} + \dots]$.}
\label{tab:anisotropic_disp}
\renewcommand{\arraystretch}{1.15}
\begin{tabular}{l c c c c c c}
\toprule
Atom & $U_{11}$ & $U_{22}$ & $U_{33}$ & $U_{23}$ & $U_{13}$ & $U_{12}$ \\
\midrule
Ce01 & 9.5(3) & 9.5(3) & 5.6(3) & 0 & 0 & 0 \\
As02 & 10.4(3) & 10.4(3) & 5.8(4) & 0 & 0 & 0 \\
As03 & 20.3(4) & 20.3(4) & 7.4(4) & 0 & 0 & 0 \\
Ni04 & 16.2(4) & 16.2(4) & 7.3(5) & 0 & 0 & 0 \\
Ni05 & 16.6(4) & 16.6(4) & 9.4(5) & 0 & 0 & 0 \\
\bottomrule
\end{tabular}
\end{table}

\newpage




\end{appendices}



\end{document}